\documentclass{rmaa}
\usepackage{paralist}

\title{revised diagnostic diagrams for planetary
nebulae}

\author{H. Riesgo and J. A. L\'opez
  \affil{Instituto de Astronom\'{\i}a, UNAM, Campus Ensenada} }

\fulladdresses{
\item Hortensia Riesgo and Jos\'e Alberto L\'opez: Instituto de
Astronom\'{\i}a, UNAM - Campus Ensenada, Apdo. Postal 877, Ensenada, B. C. 22800,
 M\'exico.   (hriesgo@astrosen.unam.mx, jal@astrosen.unam.mx)
}

\shortauthor{Riesgo \& L\'opez}
\shorttitle{Diagnostic diagrams for PNe}

\SetVolume{0} \SetFirstPage{1} \SetYear{2006}
\ReceivedDate{000} 
\AcceptedDate{000}

\resumen{Se presentan diagramas de diagn\'ostico de densidad electr\'onica -- excitaci\'on 
para una muestra de 613 nebulosas planetarias. En base a esta extensa muestra
se definen nuevos l\'{\i}mites estad\'{\i}sticos para la distribuci\'on de nebulosas
planetarias en los planos log [H$\alpha$/{[S\,\sc{ii}]}] {\textit vs} 
log [H$\alpha$/{[N\,\sc{ii}]}], 
log [H$\alpha$/{[S\,\sc{ii}]}] {\textit vs} {[S\,\sc{ii}]} $\lambda$$\lambda$6717/6731 
y log [H$\alpha$/{[N\,\sc{ii}]}] {\textit vs} {[S\,\sc{ii}]} $\lambda$$\lambda$6717/6731.
Los diagramas proveen una buena representaci\'on de los intervalos de condiciones f\'{\i}sicas,
indicados por estos cocientes de l\'{\i}neas de emisi\'on, presentes en nebulosas planetarias 
en diferentes estados de evoluci\'on.}

\abstract{Diagnostic diagrams of electron density -- excitation for a sample of
613 planetary nebulae are presented. The present extensive sample allows the definition of
new statistical limits for the distribution of planetary nebulae in the
log [H$\alpha$/{[S\,\sc{ii}]}] {\textit vs} log [H$\alpha$/{[N\,\sc{ii}]}], 
log [H$\alpha$/{[S\,\sc{ii}]}] {\textit vs} {[S\,\sc{ii}]} $\lambda$$\lambda$6717/6731 
and log [H$\alpha$/{[N\,\sc{ii}]}] {\textit vs} {[S\,\sc{ii}]} $\lambda$$\lambda$6717/6731 planes.
The diagrams provide a good representation of the ranges of physical conditions, indicated by
these emission line ratios, present in planetary nebulae during different evolutionary stages.}


\begin{document}

\maketitle

\section{Introduction}
\label{sec:Introduction}

Plots of emission line ratios provide useful diagnostics of the physical conditions and 
identification of classes of ionized gaseous nebulae. 
Electron density -- excitation diagrams were introduced by Sabbadin, Minello and
Bianchini (1977), SMB77 hereafter, with the specific purpose of clarifying the nature of the
nebula S\,176. These diagrams compare the relative line intensities of the
H$\alpha$/{[N\,\sc{ii}]}, {[S\,\sc{ii}]} $\lambda$$\lambda$6717/6731 and
H$\alpha$/{[S\,\sc{ii}]} ratios observed in supernova remnants, planetary
nebulae (PNe) and H\,II regions. These nebulae are expected to occupy distinct
regions in these diagrams as a consequence of the different physical processes operating
within them. For example, supernova remnants are mainly shock excited nebulae where low 
H$\alpha$/{[N\,\sc{ii}]} and H$\alpha$/{[S\,\sc{ii}]} ratios are expected across a
range of electron densities. H II regions are photoionized nebulae that show a 
restricted range of excitation conditions and electron densities in their generally diluted 
environments. Meanwhile, planetary nebulae show a wide range of excitation conditions and electron 
densities depending upon their evolutionary stage, the possible presence of collimated outflows
and shocks and the progenitor mass that may influence the observed H$\alpha$/{[N\,\sc{ii}]} ratio.

Recently, Phillips (2004) has discussed the 
distribution and significance of other relevant emission line ratios in PNe. 
Emission line ratio plots have also been used to distinguish planetary 
from symbiotic nebulae (e.g. Guti\'errez Moreno, Moreno \& Cort\'ez, 1995). 
The SMB77 diagrams have proven to be a useful diagnostic tool
for PNe and have been used widely in a number of works (e.g. Barral et al. 1982; L\'opez and
Meaburn 1983; Garc\'{\i}a Lario et al. 1991; Tajitsu et al. 1999). 
In the case of PNe the data in these diagrams often fall outside the original 
boundaries fit by SMB77, most likley as a consquence of the limited sample available then to set
the diagram limits. The information 
used originally by SMB77 to define the parameter space for the PNe was 
obtained from the limited compilation made by Kaler (1976), where relative emission line 
intensities observed in planetary and diffuse nebulae are listed from different sources. 
The estimated available number of PNe in that catalogue with the required
information is of only about 40 objects. The information for H II regions in SMB77 was also obtained
from that same source, whereas the data for Supernova remnants from Daltabuit et al. (1976),
Sabbadin and D'Odorico (1976) and Dopita (1976). 

Here we have taken advantage of line
intensities listed in the Strasbourg Catalogue of Galactic Planetary Nebulae,
part II, SCGPN II hereafter, (Acker et al. 1992) to explore the empirical location
of PNe in the SMB77 diagrams, but now based upon a much larger and homogeneous sample.

\section{The Sample}
\label{sec:Sample}
In order to update the parameter intervals of the
log [H$\alpha$/{[N\,\sc{ii}]}], {[S\,\sc{ii}]} $\lambda$$\lambda$6717/6731 and
log [H$\alpha$/{[S\,\sc{ii}]}] line ratios for PNe, we have compiled a database with 613 
objects from the SCGPN II that provide the line intensities required 
in the diagrams.


The H$\alpha$/{[N\,\sc{ii}]} ratio involves both [N~II] lines, 
namely $\lambda 6548 ~\&~ \lambda 6584$, whereas the
SCGPN II provides only data for the stronger $\lambda 6584$ line. We have
included the contribution from the $\lambda 6548$ line by assuming a typical
ratio $\lambda 6584 / \lambda 6548 = 2.94$. 

Figures 1, 2 and 3 show the distribution of the present dataset compared with the original 
zones for PNe, H\,II Regions and SNR as proposed by SMB77. The present
sample requires an expansion of the original axes limits.
Figure 1 shows the log [H$\alpha$/{[S\,\sc{ii}]}] {\textit vs} 
log [H$\alpha$/{[N\,\sc{ii}]}] intensity ratios; Figure 2 the log [H$\alpha$/{[S\,\sc{ii}]}]
{\textit vs} {[S\,\sc{ii}]} $\lambda$$\lambda$6717/6731 ratios and Figure 3 the 
log [H$\alpha$/{[N\,\sc{ii}]}] {\textit vs} {[S\,\sc{ii}]} $\lambda$$\lambda$6717/6731
ratios. In Figures 2 and 3 we have added dashed horizontal lines that indicate the range for 
which the {[S\,\sc{ii}]} intensity ratio is a reliable indicator of the electron density 
 (Osterbrok 1974). The corresponding
electron density scales are included on the right axes of these figures. 
Figure 1 and 4 contain the 613 objects in our sample; Figures 2 and 3 and 5 and 6, that involve 
the {[S\,\sc{ii}]} $\lambda$$\lambda$6717/6731 ratio contain only 550 objects 
for there were 52 objects with only one [S~II] line listed in the SCGPN II
which precluded obtaining the {[S\,\sc{ii}]} line ratio. Likewise, eleven other objects were
also discarded in those diagrams for having an unlikely large {[S\,\sc{ii}]} $\lambda$$\lambda$6717/6731 $>$ 2.

\begin{figure}[!ht] 
\includegraphics[width=\columnwidth]{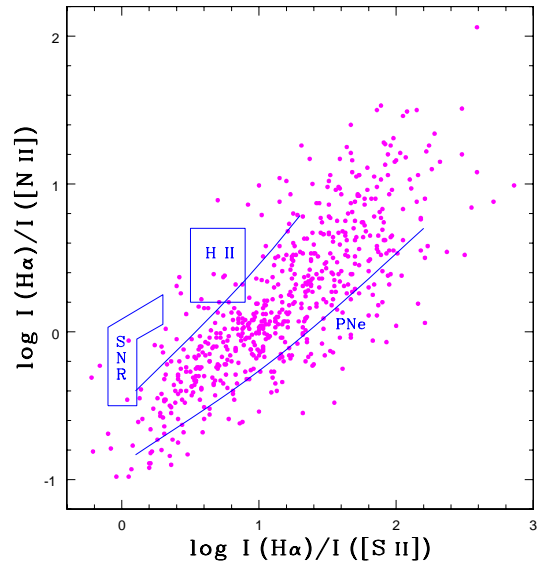}
\caption{The database used in this work is plotted in the
log [H$\alpha$/{[S\,\sc{ii}]}] {\textit vs}
log [H$\alpha$/{[N\,\sc{ii}]}] plane. The original zones for
SNR, H II regions and PNe proposed by SMB77 are indicated.}
\label{fig:1}
\end{figure}

\begin{figure}[!ht] 
\includegraphics[width=\columnwidth]{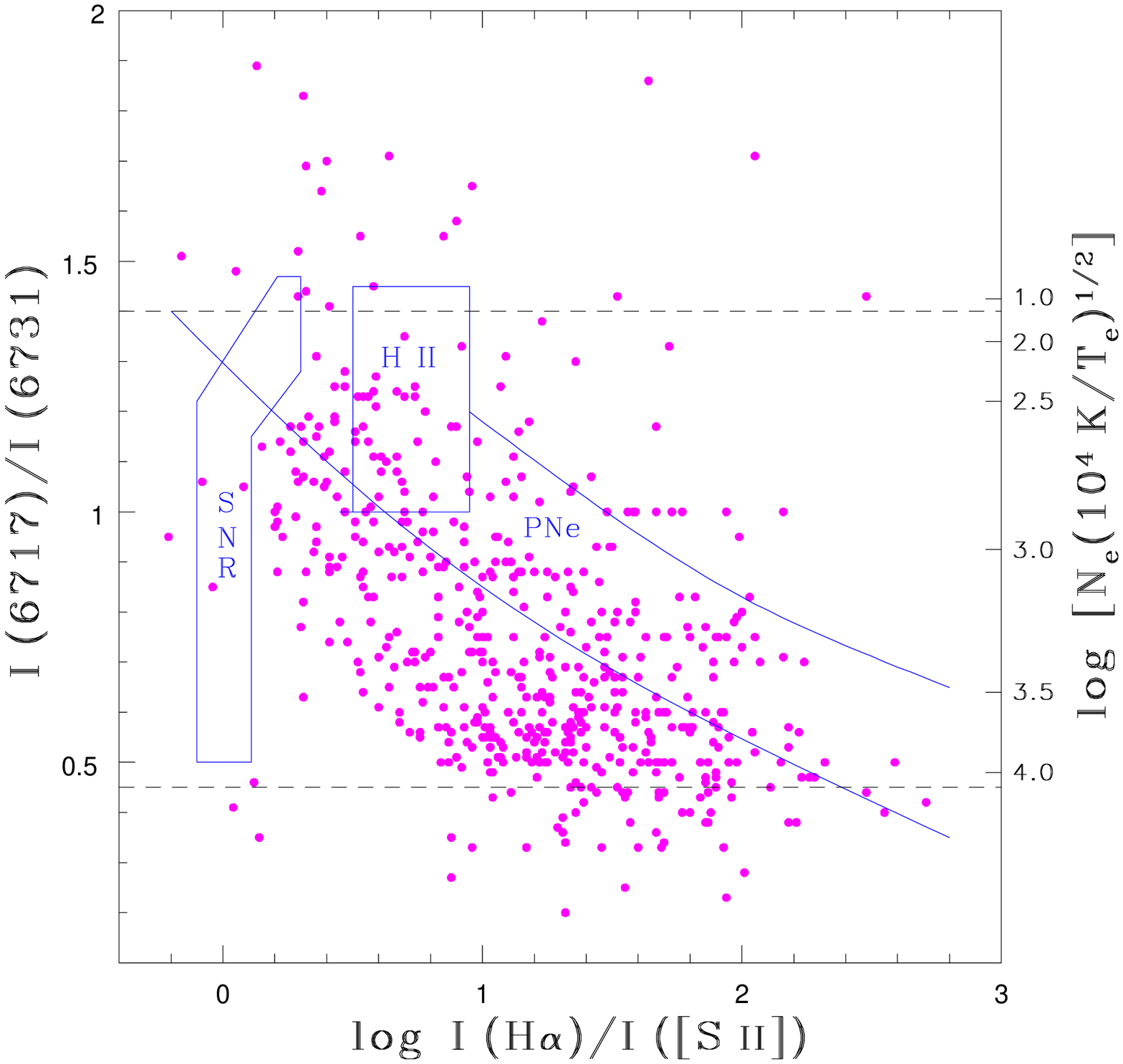}
\caption{As in Fig. 1, but for the log [H$\alpha$/{[S\,\sc{ii}]}] {\textit vs} 
{[S\,\sc{ii}]} $\lambda$$\lambda$6717/6731 plane. The horizontal dashed lines
indicate the range of values for which the {[S\,\sc{ii}]} $\lambda$$\lambda$6717/6731
ratio is a valid indicator of the electron density.} 
\label{fig:2}
\end{figure}

\begin{figure}[!ht] 
\includegraphics[width=\columnwidth]{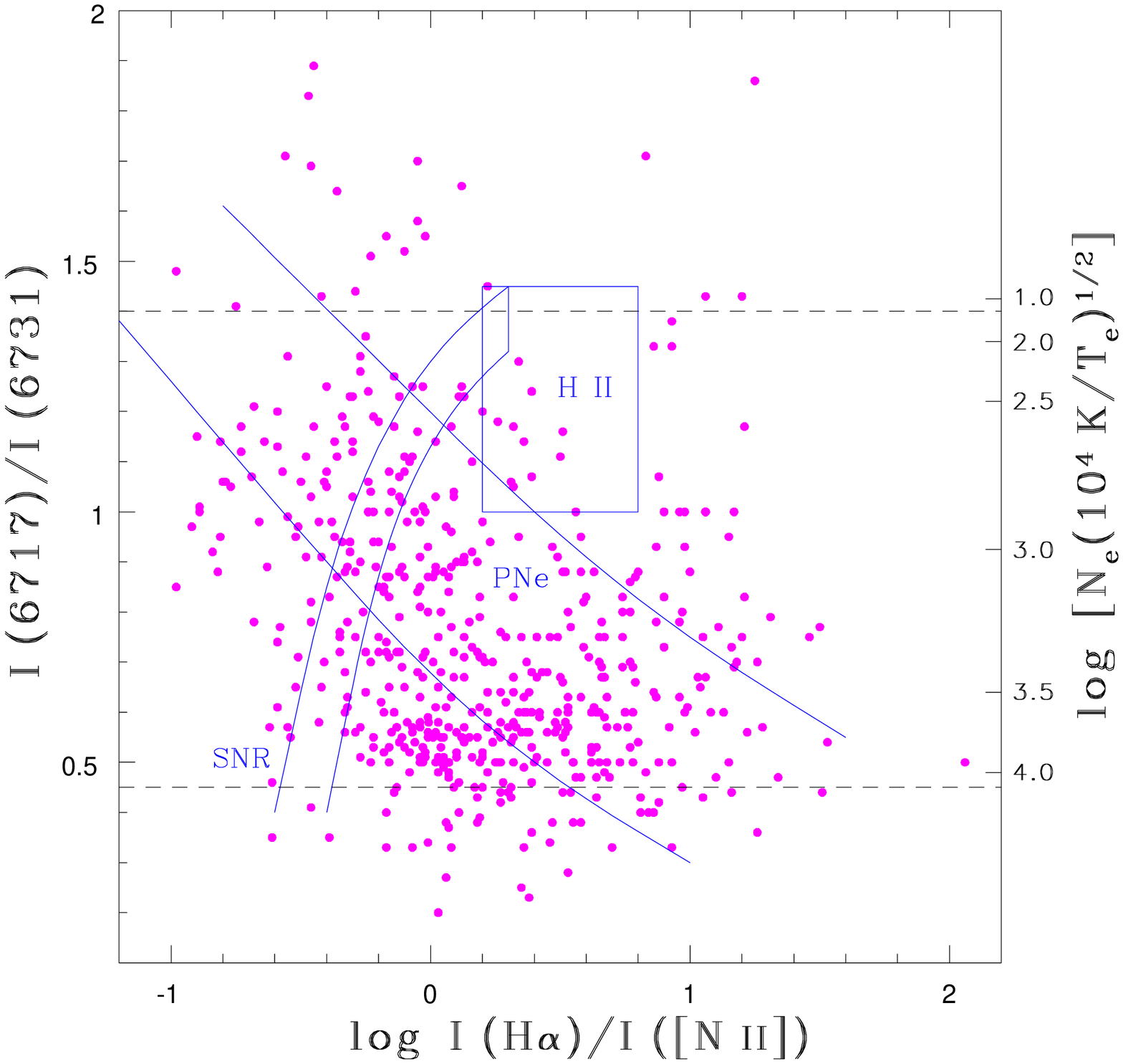}
\caption{As in Fig. 1, but for the log [H$\alpha$/{[N\,\sc{ii}]}] {\textit vs}
{[S\,\sc{ii}]} $\lambda$$\lambda$6717/6731 plane. The horizontal dashed lines
indicate the range of values for which the {[S\,\sc{ii}]} $\lambda$$\lambda$6717/6731
ratio is a valid indicator of the electron density.} 
\label{fig:3}
\end{figure}

\begin{figure}[!ht] 
\includegraphics[width=\columnwidth]{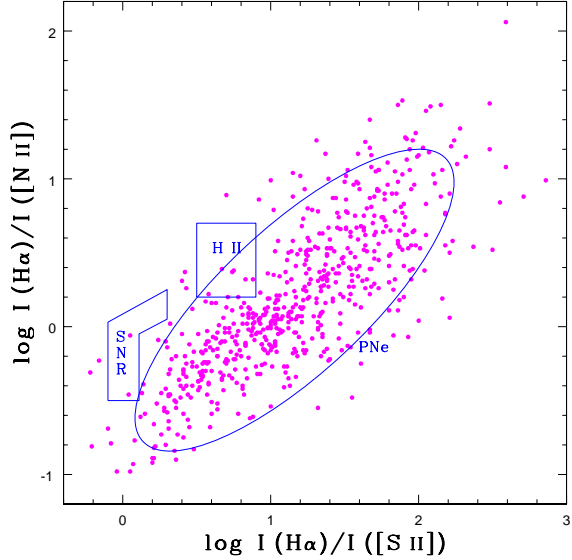}
\caption{Diagnostic diagram for the log [H$\alpha$/{[S\,\sc{ii}]}] intensity ratio 
{\textit vs} log [H$\alpha$/{[N\,\sc{ii}]}] ratio with the new limits defined by
a density ellipse of probability 0.85. The correlation coefficient is 0.79}
\label{fig:4}
\end{figure}

\begin{figure}[!ht] 
\includegraphics[width=\columnwidth]{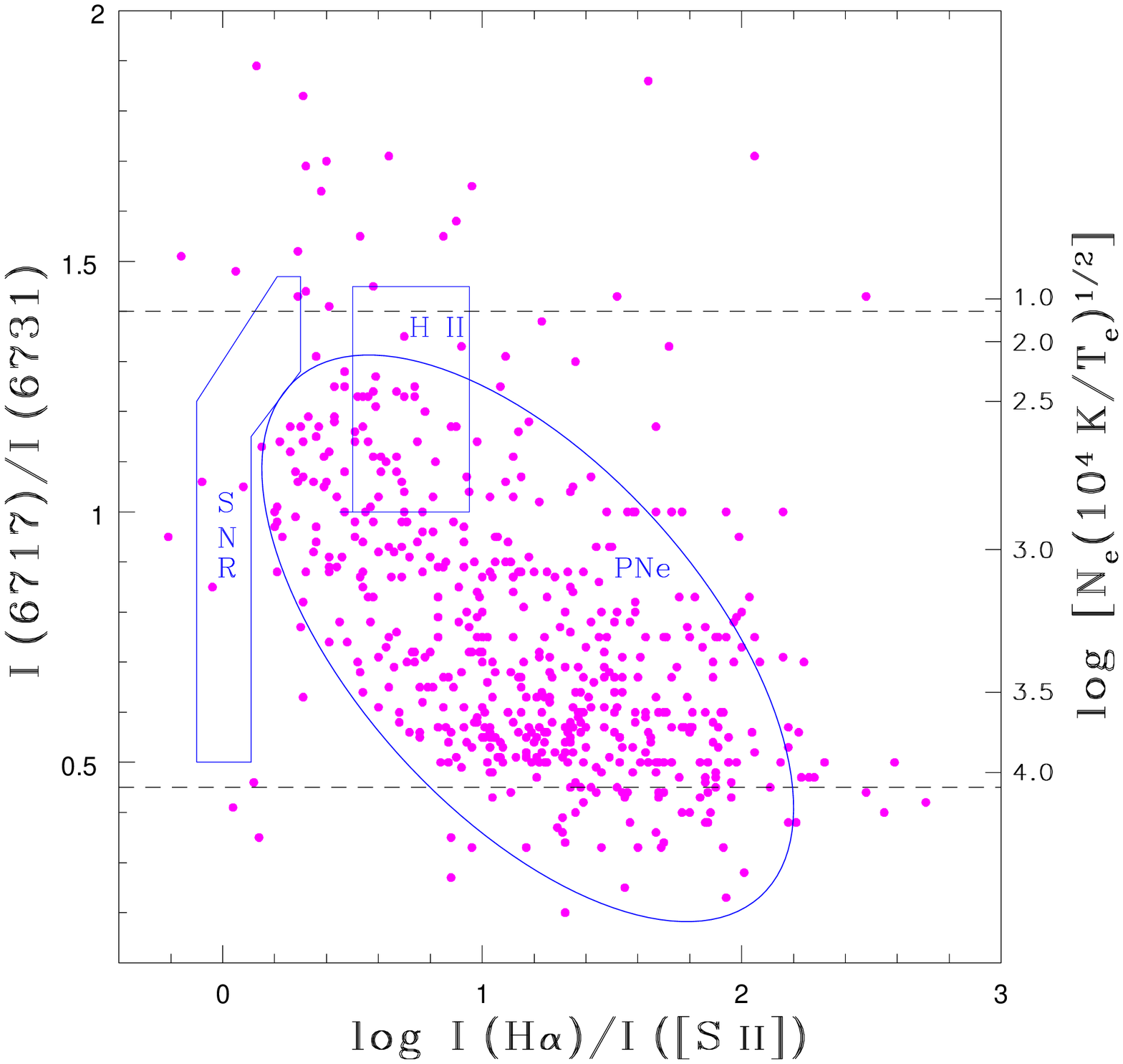}
\caption{As in Fig. 4, but for the log [H$\alpha$/{[S\,\sc{ii}]}] {\textit vs} 
{[S\,\sc{ii}]} $\lambda$$\lambda$6717/6731 plane. The correlation coefficient is -0.58 }
\label{fig:5}
\end{figure}

\begin{figure}[!ht] 
\includegraphics[width=\columnwidth]{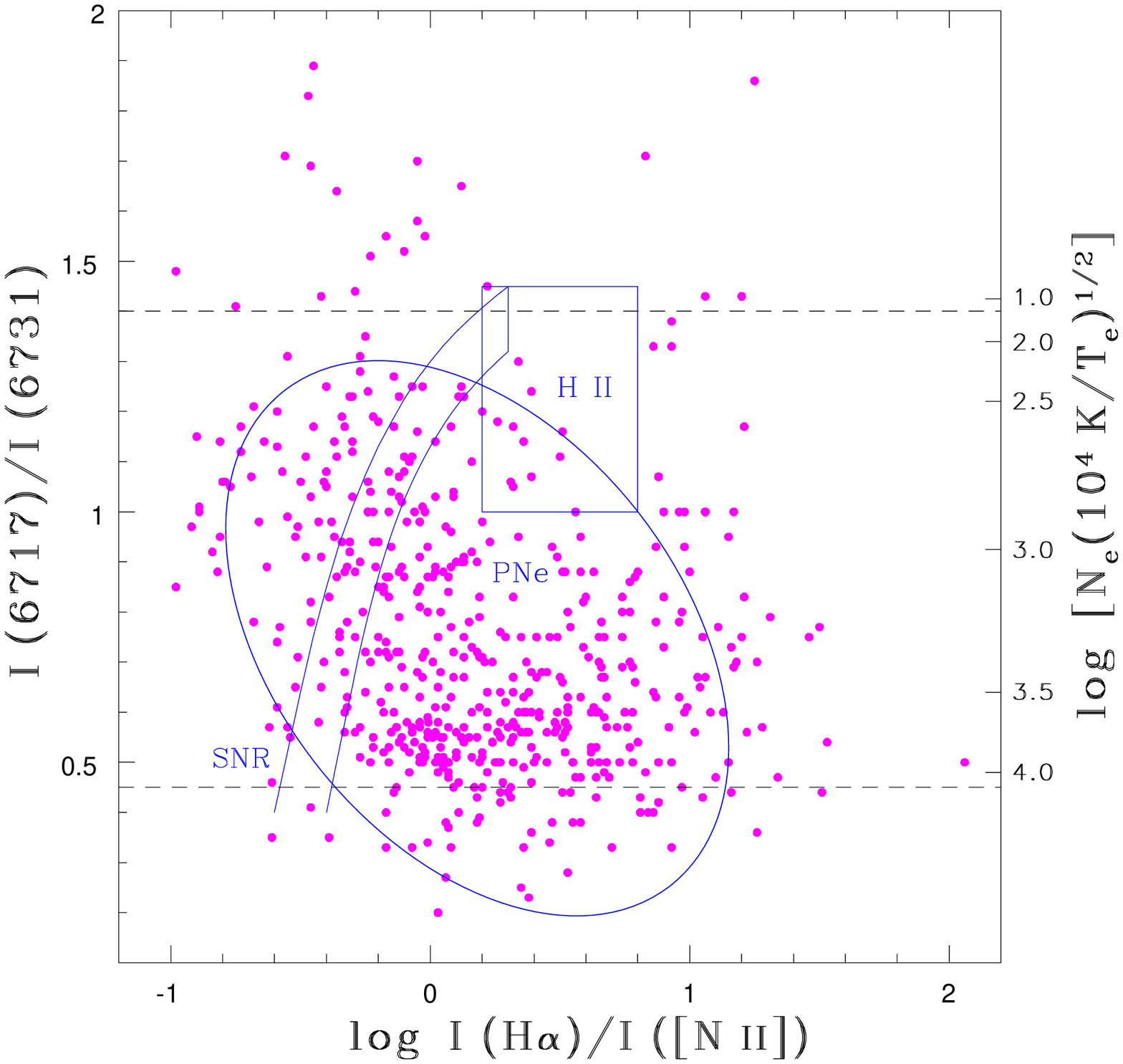}
\caption{As in Fig. 4, but for the log [H$\alpha$/{[N\,\sc{ii}]}] {\textit vs}
{[S\,\sc{ii}]} $\lambda$$\lambda$6717/6731 plane. The correlation coefficent is -0.37}
\label{fig:6}
\end{figure}


\begin{figure}[!ht] 
\includegraphics[width=\columnwidth]{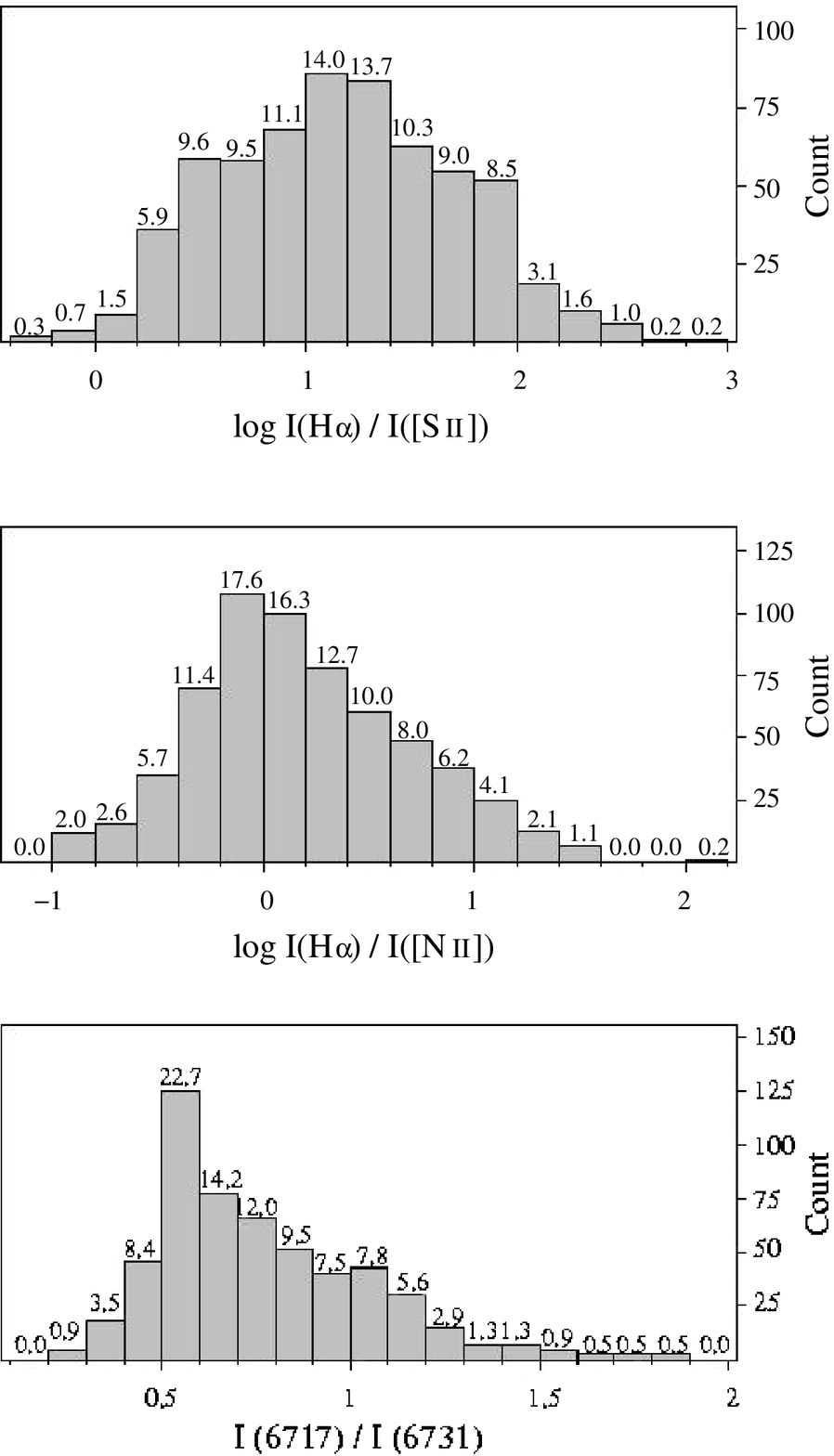}
\caption{Histograms of the distribution of the values for the corresponding line ratios. Numbers at the top
of each bar correspond to the percentage value of that column in relation to the total sample}
\label{fig:7}
\end{figure}




\section{Discussion}
It is clear from Figures 1 -- 3 that the limits and zones defined for PNe in the
original sample used by SMB77 fall short of covering substantial data regions in these diagrams. 
The present extended sample from the SCGPN II expands the previous location of PNe in the 
diagrams, providing a more accurate representation of their zones of influence.

New statistical limits therefore have now been set to represent the 
location of PNe in the SMB77 diagrams by means of density ellipsoid probabilities.
Figures 4, 5 and 6 show the new zones defined by ellipses of probablity 0.85 for the 
corresponding emission line ratios for PNe. The ellipse in the diagrams represents both 
the density contour and confidence region of finding an object within this zone with an 
85 \% confidence level. The density ellipse is also a good graphical indicator of the 
correlation between the two variables involved. The ellipse collapses across its minor axis, 
i.e., becomes thinner, 
as the correlation betwen the two variables increases and approaches unity ($\pm 1$) and expands 
or reduces 
its eccentricity as the correlation weakens. The density ellipses in Figures 4, 5 \& 6 
have all been set for a 0.85 probabilty, however the correlation coefficients differ among them, 
with values of 0.79, -0.58 and -0.37, respectively. Thus, the correlation between 
log [H$\alpha$/{[S\,\sc{ii}]}] {\textit vs} log [H$\alpha$/{[N\,\sc{ii}]}] is strongest 
whereas that between log [H$\alpha$/{[N\,\sc{ii}]}] {\textit vs} {[S\,\sc{ii}]} $\lambda$$\lambda$6717/6731
is weakest. The correlation is related to the scatter
of the points in the diagrams and this scatter is a combination of real dispersion due to different
physical conditions, as those expected from different evolutionary stages of
a planetary nebula, different conditions within a nebula (e.g. core, rim or halo) and in some 
cases, also probable observational uncertainties or measuring errors from the spectroscopic SCGPN II 
survey. Phillips (2004) has discussed a number of factors that potentially affect 
such samples, such as the aperture (slit) size and its location over the nebula.

The samples for each diagram follow the trends apparent in the
original SMB77 diagrams, though now they cover zones where the concentration and correlation of 
the line ratios provide a much closer representation of 
the true location of PNe in these diagnostic diagrams. For example, the ellipses in Figures 5 and 6 show  
a clear tendency towards larger electron density values as compared to the corresponding original 
areas indicated in Figures 3 and 4. These values are in good accord with the electron densities 
derived from other, smaller, samples such as in Stanghellini and Kaler (1989).

The diagrams provide direct and useful information on general physical conditions in PNe. For example,
diluted shells with low electron density and low to moderate excitation conditions, e.g. 
$0.5 <$log [H$\alpha$/{[S\,\sc{ii}]}] $<1$ (see Fig. 5) will overlap the location of H II regions. 
These objects are likely large, evolved PNe. The likely presence of shock contributions in the
dominantly photoionized environment of PNe is hinted by objects located in Fig. 4 in the 
log [H$\alpha$/{[N\,\sc{ii}]}]$<0$ and log [H$\alpha$/{[S\,\sc{ii}]}] $< 0.4$ region. Objects with
high electron density and high H$\alpha$ emissivity in Figs. 5 and 6 point towards compact, young PNe.
For PNe for which limited information is available, their location on the diagrams usually provides
insight into their general (or peculiar) characteristics and evolutionary stage.

Figure 7 shows the histograms of the distribution of values for the corresponding line
ratios. The numbers at the top of each bar indicate the percentage of objects in the sample associated 
with that value in relation to the entire sample. The line ratios of PNe are concentrated around a 
value of 1.2 
for log [H$\alpha$/{[S\,\sc{ii}]}] with a fairly even distribution on either side 
from the peak. For the log [H$\alpha$/{[N\,\sc{ii}]}] line ratio its values
concentrate around 0.0 but falling rapidly to negative values. Finally, the distribution of 
values for the {[S\,\sc{ii}] $\lambda$$\lambda$6717/6731} line ratio is markedly asymmetric, peaking 
sharply at a value of 0.6, with a smooth drop towards larger values and an abrupt fall on 
the other side (high density). These values can be considered as representative for 
the most common or typical conditions in galactic PNe.

\section{Conclusions}
\label{sec:Conclusions}

Values of H$\alpha$/{[N\,\sc{ii}]}, {[S\,\sc{ii}]} $\lambda$$\lambda$6717/6731 and
H$\alpha$/{[S\,\sc{ii}]} for 613 planetary nebulae have been derived from the Strasbourg
Catalogue of Planetary Nebulae, part II (Acker et al. 1992) and plotted in the Sabbadin,
Minello \& Bianchini (1977) diagnostics diagrams. The current sample substantially expands 
the sample used in the original diagrams and allows a redefinition of the parameter space of PNe 
in the 
log [H$\alpha$/{[S\,\sc{ii}]}] {\textit vs} log [H$\alpha$/{[N\,\sc{ii}]}], 
log [H$\alpha$/{[S\,\sc{ii}]}] {\textit vs} {[S\,\sc{ii}]} $\lambda$$\lambda$6717/6731 
and log [H$\alpha$/{[N\,\sc{ii}]}] {\textit vs} {[S\,\sc{ii}]} $\lambda$$\lambda$6717/6731 planes.
Accordingly, new statistical limits have been set for the location of PNe in these diagrams using density
ellipses of probability 0.85. We find a good, 0.79 correlation for the log [H$\alpha$/{[S\,\sc{ii}]}] 
{\textit vs} log [H$\alpha$/{[N\,\sc{ii}]}] intensity ratio; a modest -0.58 correlation for the 
log [H$\alpha$/{[S\,\sc{ii}]}] {\textit vs} log [H$\alpha$/{[N\,\sc{ii}]}] intensity ratio 
and a poor, -0.37 correlation in the case of 
log [H$\alpha$/{[N\,\sc{ii}]}] {\textit vs} {[S\,\sc{ii}]} $\lambda$$\lambda$6717/6731.
 From our study we obtain the statistical 
distribution of emission line ratios expected in PNe. 
These upgraded diagnostic diagrams provide a better representation of
the ranges of physical conditions indicated from the emission line ratios and allows the identification
and study of groups or individual PNe of particular interest in specific regions of the diagrams.

Finally, in relation to the parameter space for the H II and SNR regions, the data sources 
in SMB77 for these two cases were restricted to a limited number
of objects, as in the original case for PNe. For the HII regions they used the compilation made by 
Kaler (1976). For the supernovae the data is based on 17 galactic remnants. It would be clearly useful 
to revise and update the parameter space for the H II and SNR regions to assess the reliability of their 
distributions in the revised diagrams. This is presently out of the scope of the present paper but 
worth pursuing as a natural extension of this work. 

Templates in poscript or pdf format of the revised diagrams, Figures 4, 5 and 6, are available 
from J. A. L\'opez.

\acknowledgements

HR acknowledges a posgraduate scholarship from CONACYT. JAL acknowledges
continued support from DGAPA-UNAM and CONACYT through grants IN112103 and
43121, respectively. We thank M. Richer and W. Steffen for their comments on this work and the 
anonymous referee for helping us to improve the presentation of the paper.

\end{document}